\documentclass{arxiv2}

\begin{document}

\catchline{}{}{}{}{}
%%%%%%%%%%%%%%%%%%%%%%%%%%%%%%%%%%%%%%%%%%%%%%%%%%%%%%%%%%%%%%%%%%%

\title{Order in the mass spectrum of elementary particles 
}

\author{T. A. Mir$^{*}$ and G. N. Shah }

\address{Nuclear Research Laboratory, Bhabha Atomic Research Centre, \\
Zakura, Srinagar - 190 006, Jammu and Kashmir, India\\
$^{*}$taarik.mir@gmail.com}

\maketitle

%\pub{Received (Day Month Year)}{Revised (Day Month Year)}

\section{Abstract}
The study on the linkage of elementary particle mass differences with pion-muon mass difference is explored further. In the present study we show this linkage to be equally true for the mass differences amongst the members of SU(3) hadron multiplets and the hadrons belonging to multiplets having different spin and parity characteristics. This reinforces our contention that inter particle mass excitations are quantized as integral multiples of basic mass unit of 29.318 MeV and thereby indicates the fundamental nature of pion-muon mass difference in the elementary particle mass distribution.
\section{Keywords}

Hadrons; SU(3) multiplets; mass quantization.
%\end{abstract}

\ccode{PACS Nos.: 12.10.Kt, 12.40.Yx, 13.40.Dk}

\section{Introduction}	
Although the standard model is in excellent agreement with the rich experimental data, it leaves many questions unanswered. The observed mass spectrum of the elementary particles is one of the fundamental questions that has so far defied any reasonable explanation in the standard model\cite{Christianto}$^{,}$ \cite{Fritzsch}. The distribution of the elementary particle masses is quite bizzare and is spread from a fraction of eV's for neutrinos to hundreds of GeV's for the top quark. Apart from few patterns based on SU(3) symmetry that led to the Gell-Mann Okubo\cite{Okubo} and Coleman-Glashow\cite{Coleman} formulae interelating the masses of pseudoscalar mesons and octet baryons respectively, the standard model has not revealed any general order in the elementary mass spectrum. On the other hand the associated physical mass is the best known and most fundamental characteristic property of the elementary particles, the recognition of empirical systematics and of any general regularities in the mass spectrum of these particles, irrespective of their nature or scheme of classiﬁcation is of tremendous importance in understanding the intricacies of mass spectrum of elementary particles. The lowest mass states, as a general rule in physical systems, are considered to be the building blocks of more complex systems and hence in some sense the most fundamental. The most stable and least massive particles i.e. electron, muon and pion to which other particles decay after a transient existence, are the natural candidiates to search for a systematic regularity within the mass spectrum of elementary particles\cite{Mac Gregor3}. Empirical and theoritical investigations based on experimental data reveal the electron, muon and the pion to serve as basic units for exploring the discrete nature of the mass distribution of elementary particles\cite{Shah}$^{-}$\cite{Akers}. To search for an order, we perform a specific systematic  analysis of the mass spectrum of elementary particles and reveal that mass differences among particles when arranged in the ascending order of mass have a general tendency to be close integral/half integral multiple of mass difference between a neutral pion and a muon i.e. 29.318 MeV. The mass differences between unstable leptons and between baryons were shown to be quantized as integral multiples of this basic unit of 29.318\cite{Shah}. In the present study, we evaluate the applicability of this result to the SU(3) hadron multiplets and to neutral hadrons. We reveal that mass unit of about 29.318 MeV is a determining factor for the distribution of mass of elementary particles by showing that 29.318 MeV integral multiplicity of mass differences to be valid for these hadrons. This reinforces our earlier result that elementary particles do not occur randomly and are linked through the mass difference between first two massive elementary particles i.e. a muon and a neutral pion.   
\section{Data Analysis and Results}
The database for the present study is the latest version of the Particle Data Group listings\cite{Yao}. Here we investigate relationship the pion-muon mass difference has with the mass structure of the 1) hadrons which are classified into multiplets on the basis of SU(3) symmetry and 2) neutral hadrons.
\subsection{Baryon Multiplets}
\subsubsection{Baryon Octet}The masses of the baryon octet members with spin J and parity P such that $J^{P}$=$\frac{1}{2}$$^{-}$ are: $m_{p}$=938.27203 MeV, $m_{n}$=939.56536 MeV, $m_{\Lambda^{0}}$=1115.683 MeV, $m_{\Sigma^{+}}$=1189.37 MeV, $m_{\Sigma^{-}}$=1197.449, $m_{\Xi^{0}}$=1314.83 MeV and $m_{\Xi^{-}}$=1321.31 MeV. The successive mass differences are tabulated in Column 1 of Table 1 with numerical value in MeVs given in Column 2. The small mass differences between the different members of an isospin charge multiplet are known to arise from the electromagnetic interaction\cite{Perkins}.
However, the masses of the members of different isospin multiplets differ considerably. Column 4 shows the integral multiples of 29.318 MeV that are close to the observed mass difference between successive members of the octet. The integers being shown in Column 3. The deviations of the observed value from the closest integral multiple of 29.318 MeV are given in Column 5. It is observed that the mass difference between $\Lambda^{0}$ and $n$ i.e. 176.118 MeV differs from the nearest predicted value of 175.908 MeV by only 0.21 MeV. Same is true of the mass difference i.e. 117.381 MeV between the particles $\Xi^{0}$ and $\Sigma^{-}$ which differs from the predicted value of 117.272 MeV by only 0.109 MeV. However, observed mass interval of $\Sigma^{+}$ and $\Lambda^{0}$ differs from the predicted value by about 14.264 MeV. 
Interestingly, this large value turns out to be half integral multiple of the mass difference between a $\pi^{0}$ and a $\mu^{-}$. As can be clearly seen from the row 3 of Table 1, the observed mass difference between $\Sigma^{+}$ and $\Lambda^{0}$ i.e. 73.69 differs from the half integral ($\frac{5}{2}$) multiple of pion and muon mass difference by only 0.39 MeV. The maximum mass splitting within the baryon octet i.e. mass difference of 383.037 MeV, between  the heaviest member $\Xi^{-}$ and the lightest baryon $p$ is close integral multiple of 29.318 MeV, differing from the predicted value by only 1.904 MeV. It may be pointed out that 29.318 MeV multiplicity also holds for the mass intervals among any of the octet members\cite{Shah}. Clearly the 29.318 MeV multiplicity holds with great precision for the baryon octet members.   
\begin{table}[h]
\tbl{The observed baryon octet mass intervals as integral multiple of 29.318 MeV}
{\begin{tabular}{@{}lllllll@{}} \toprule
Particles & Mass Difference & Integer & $N$$\times$29.318  & Obsd - Expd  \\
& (MeV) & $N$  & (MeV) & (MeV)\\ \colrule

$\Lambda^{0}$ - $n$\hphantom{00} & 176.118 & 6 & 175.908 & 0.21 \\ 
\\
$\Sigma^{+}$ - $\Lambda^{0}$\hphantom{00} & 73.687 & 3 & 87.954 & 14.267 \\
& & ($\frac{5}{2}$) & 73.295 & 0.392 \\ 
  & \\ 
$\Xi^{0}$ - $\Sigma^{-}$\hphantom{00} & 117.381 & 4 & 117.272 & 0.109 \\
\\
$\Xi^{-}$ - p\hphantom{00} & 383.038 & 13 & 381.134 & 1.904 \\
\\
\botrule
\end{tabular} \label{ta1}}
\end{table}

\subsubsection{Baryon Decuplet} The analysis for the baryon decuplet members with $J^{P}$=$\frac{3 }{2}$$^{-}$ is detailed in Table 2. It may be pointed out that while all the members of the baryon octet are non-resonant states, for the baryon decuplet all the members execpt for the $\Omega^{-}$ are resonances. Since the Particle Data Group reports an average mass for the four charged states of the $\Delta$ baryons and individual masses for the different charge states of $\Sigma^{*}$ and  $\Xi^{*}$ baryons, we cosider the average masses of each isospin multiplet which are as follows: $m_{\Delta}$=1232 MeV, $m_{\Sigma^{*}}$=1384.56 MeV,  $m_{\Xi^{*}}$=1533.4 MeV, and $m_{\Omega^{-}}$=1672.45 MeV.
The first three mass differences in the Table 2 are those between the successive members of the decuplet. The equal spacing rule for the SU(3) decuplet predicts masses of successive isospin multiplets to be equidistant\cite{Oh} i.e. $m_{\Sigma^{*}}$ - $m_{\Delta}$ = $m_{\Xi^{*}}$ - $m_{\Sigma^{*}}$ = $m_{\Omega^{-}}$ - $m_{\Xi^{*}}$. However, this rule is not strictly obeyed in the decuplet, since the mass separations are not exactly same as evident from Table 2. Further, although the mass spacing among the successive decuplet members deviate from the closest integral (5) multiples of 29.318 MeV by 5.97, 2.25 and 7.54 MeV respectively, it is important to note that the average mass spacing among successive members i.e. 146.816 MeV is very close to 146.59 MeV, a value obtained on integral (5) multiplication of 29.318 MeV. The difference between the observed and predicted values being 0.226 MeV only. This may be compared with 140 MeV pionic mass interval\cite{Akers} among the decuplet members which deviates from the average mass spacing by about 6.816 MeV\cite{Mac Gregor4}. The $\Omega^{-}$ - $\Delta$ is the  difference between the mass of lightest resonance member $\Delta$ and the heaviest non-resonant member $\Omega^{-}$ of the decuplet. This observed mass interval of 440.45 MeV is very close to 439.77 MeV, obtained on integral (15) multiplication of the mass difference between a neutral pion and a muon. The difference between the observed and expected value being only 0.68 MeV. 
\begin{table}[h]
\tbl{The observed baryon decuplet mass intervals as integral multiple of 29.318 MeV}
{\begin{tabular}{@{}lllllll@{}} \toprule
Particles & Mass Difference & Integer & $N$$\times$29.318 & Obsd - Expd \\
& (MeV) &  $N$  & (MeV) & (MeV) \\ \colrule

$\Sigma^{*}$ - $\Delta$ \hphantom{00} & 152.56 & 5 & 146.59 & 5.97 \\ 
\\
$\Xi^{*}$ - $\Sigma^{*}$\hphantom{00} & 148.84 & 5 & 146.59 & 2.25 \\
\\ 
$\Omega^{-}$ - $\Xi^{*}$\hphantom{00} & 139.05 & 5 & 146.59 & 7.54 \\
 \\
Average decuplet mass spacing\hphantom{00} & 146.816 & 5 & 146.59 & 0.226\\
\\
$\Omega^{-}$ - $\Delta$\hphantom{00} & 440.45 & 15 & 439.77 & 0.68 \\
\\

\botrule

\end{tabular} \label{ta1}}
\end{table}

\subsection{Meson Multiplets}
\subsubsection{Pseudoscalar Meson Nonet} The mesons with spin zero and odd parity i.e. $J^{P}$ = $0^{-}$ are organized into a multiplet contaning nine states to form pseudoscalar meson nonet. These mesons have the lowest rest energy. In column 1 of Table 3, the k$^{\pm}$ - $\pi^{\pm}$, $\eta$ - k$^{\pm}$ and $\eta^{'}$ - $\eta$ are the mass difference between the successive members of the pseudoscalar meson nonet with numerical values  given in column 2. The deviation of 0.182 MeV between the observed and predicted $\eta^{'}$ - $\eta$ mass difference may be compared with that of 0.730 MeV obtained from the integral multiple (3) of the 137 MeV\cite{Mac Gregor4}. Both the observed mass intervals of $\eta$ - $\pi^{0}$ and $\eta$ - $\pi^{\pm}$ deviate from the 14 multiples of 29.318 MeV by about 2 MeV. However the difference 410.2364 MeV between the mass of $\eta$ and average pion mass of 137.27339 MeV differs by only 0.215 MeV from predicted value 410.452 obtained by the multiplication of 29.318 MeV by integer 14. Again this departure of the observed mass difference between $\eta$ and average pion mass may be comapared with the difference of 0.736 MeV bewteen the observed and that expected when mass intervals are taken as integral multiples of average pion mass i.e. 137 MeV\cite{Mac Gregor4}.  The observed mass interval between the $\eta^{'}$ and average pion mass is 820.506 MeV. This value deviates from the predicted value of 820.904 MeV obtained from the integral multiple (28) of 29.318 MeV by only 0.3973 MeV only. Whereas the predicted value 822 MeV on the basis of integral pion mass differs from the observed value by 1.4933 MeV\cite{Mac Gregor2}. The difference $\eta^{'}$ - $\pi^{0}$ is the mass difference between the lightest and heaviest member of the pseudoscalar meson nonet. As can be seen from the Table 3, the observed mass spacings are close integral multiples of the mass difference between a neutral pion and a muon. The masses for the pseudoscalar mesons taken from the Particle Data Group listings are: $m_{\pi^{0}}$=134.9766 MeV, $m_{\pi^{\pm}}$=139.57018 MeV, $m_{k^{\pm}}$=493.677 MeV, $m_{k^{0}}$= 497.648 MeV, $m_{\eta}$=547.51 MeV, $m_{\eta^{'}}$=957.78 MeV.
\begin{table}[h]
\tbl{The observed pseudoscalr meson mass intervals as integral multiple of 29.318 MeV}
{\begin{tabular}{@{}lllllll@{}} \toprule
Particles & Mass Difference & integer  & $N$$\times$29.318  & Obsd - Expd  \\
& (MeV) & $N$ &  (MeV) & (MeV)\\ \colrule
k$^{\pm}$ - $\pi^{\pm}$\hphantom{00} & 354.107 & 12 & 351.816 & 2.291 \\
\\

$\eta$ - k$^{\pm}$\hphantom{00}   & 53.833   & 2 & 58.636 & 4.803   \\
\\
$\eta^{'}$ - $\eta$\hphantom{00} &    410.27    & 14 & 410.452 & 0.182 \\
\\
$\eta$ - $\pi^{0}$\hphantom{00} &    412.5344    & 14 & 410.452 & 2.0814 \\ 
\\
$\eta$ - $\pi^{\pm}$\hphantom{00} &  407.93982    & 14 & 410.452 & 2.51218 \\ 
\\
$\eta$ - $\pi_{avg}$\hphantom{00} &  410.2364    & 14 & 410.452 & 0.215 \\ 
\\
$\eta^{'}$ - k$^{\pm}$\hphantom{00} &  464.103    & 16 & 469.088 & 4.985 \\
\\
$\eta^{'}$ - $\pi^{0}$\hphantom{00} & 822.81 & 28 & 820.904 & 1.906 \\
\\
$\eta^{'}$ - $\pi^{\pm}$\hphantom{00} & 818.20982 & 28 & 820.904 & 2.69418 \\
\\
$\eta^{'}$ - $\pi_{avg}$\hphantom{00} & 820.5067 & 28 & 820.904 & 0.3973 \\
\\ \botrule

\end{tabular} \label{ta1}}
\end{table}
\subsubsection{Vector Meson Nonet}The nine vector mesons with spin one and odd parity i.e. $J^{P}$ = $1^{-}$ form a multiplet called vector meson nonet. The analysis for the mass differences among the vector meson nonet members is detailed in Table 4. The mass difference between successive members (Column 1) of the vector meson nonet  is given in the Column 2. As is evident from the Table 4, the observed mass differences k$^{*}$ - $\rho$ and $\phi$ - $\omega$ are close integral multiples of 29.318 MeV. Although the mass difference between isospin triplet $\rho$ meson and isospin singlet $\omega$ meson 7.15 MeV is nonelectromagnetic in origin but is of the order of electromagnetic mass splitting. The deviation of 2.226 MeV of the observed $\phi$ - $\omega$ mass difference from the integral multiple of 29.318 MeV is in better agreement than that of 8.19 MeV obtained when mass interval is predicted as half integral multiple of a mass unit of about 70 MeV i.e about half the pion mass\cite{Mac Gregor2}. Since the Particle Data Group lists average mass for the isospin triplet $\rho$ states and separate masses for two charge states of k$^{*}$ meson, we consider the average mass of vector mesons for the analysis which are: $m_{k^{*}}$=893.83 MeV, $m_{\rho}$=775.5 MeV, $m_{\omega}$=782.65 MeV and $m_{\phi}$=1019.460 MeV.
\begin{table}[h]
\tbl{The observed vector meson mass intervals as integral multiple of 29.318 MeV}
{\begin{tabular}{@{}lllllll@{}} \toprule
Particles & Mass Difference & integer  & $N$$\times$29.318  & Obsd - Expd  \\
& (MeV) & $N$ &  (MeV) &  (MeV) \\ \colrule
k$^{*}$ - $\rho$\hphantom{00} & 118.33 & 4 & 117.272 & 1.058 \\
\\

$\omega$ - $\rho$\hphantom{00}   & 7.15   &  &  &   \\
\\
$\phi$ - $\omega$\hphantom{00} &    236.81    & 8 & 234.544 & 2.266 \\

\\ \botrule

\end{tabular} \label{ta1}}
\end{table}
\subsection{Inter multiplet mass intervals}
The applicabilty of 29.318 MeV multiplicity of elementary particle mass intervals extends beyond the mass differences between the successive members of a particular SU(3) multiplet and applies to mass intervals among the members of multiplets with different spin and parity characteristics. The analysis for the mass intervals between the octet and decuplet baryons is detailed in Table 5. The lightest member of decuplet i.e $\Delta$(1232) resonance is the first excited state of the proton and the observed mass interval between the two is accounted by the hyperfine splitting due to colour magnetic interaction among the quarks\cite{Perkins}. However, from Table 5 it is seen that the observed mass interval of 293.728 MeV between $\Delta$(1232) and proton is very close to 293.180 MeV, a value obtained on integral multiplication of 29.318 by 10. The difference between the two values being only 0.548 MeV. The relation  of
$\Delta$(1232) and proton is expected as $\Delta$ baryons are considered to be the excited states of nucleon but there is no relation between $\Delta$(1232) and $\Lambda^{0}$. However, from our analysis it follows that the observed mass interval between the two 116.317 MeV differs from 117.272 MeV, the closest integral (4) multiple of 29.318 MeV by 0.955 MeV only. Thus $\Delta$(1232) can be obtained by taking four excitations of 29.318 MeV from the $\Lambda^{0}$. Similarly the observed mass inteval of 556.767 bewteen two unrelated baryons $\Omega^{-}$ and $\Lambda^{0}$ deviates from 557.042 MeV, a value obtained as 19 multiples of pion-muon mass difference by only 0.275 MeV. Further, the observed mass difference 351.14 MeV between the heaviest member of baryon octet $\Xi^{-}$ and that of baryon decuplet i.e. $\Omega^{-}$ the only non-resonant member, differs from 351.816 MeV, a value obtained on integral (12) multiplication of 29.318 MeV, by only 0.676 MeV. The detailed anlysis of the mass differences among the pseudoscalar and vector mesons is given in Table 6. As seen from the Table 6 the 29.318 MeV mass interval multiplicity is valid for mesons also. 
\begin{table}[h]
\tbl{The observed mass intervals between octet and decuplet barayons as integral multiple of 29.318 MeV}
{\begin{tabular}{@{}lllllll@{}} \toprule
Particles & Mass Difference & Integer & $N$$\times$29.318 & Obsd - Expd\\
& (MeV) & $N$  & (MeV) & (MeV)\\ \colrule

$\Delta$(1232) - p\hphantom{00} & 293.728 & 10 & 293.180 & 0.548 \\ 
\\

$\Delta$(1232) - $\Lambda^{0}$\hphantom{00} & 116.317 & 4 & 117.272 & 0.955 \\ 
\\
$\Delta$(1232) - $\Sigma^{-}$\hphantom{00} & 34.551 & 1 & 29.318 & 5.233 \\ 
\\
$\Delta$(1232) - $\Xi^{-}$\hphantom{00}& 89.31 & 3 & 87.954 & 1.356 \\
\\
$\Sigma^{*+}$(1382.8) - n\hphantom{00} & 443.15 & 15 & 439.77 & 3.38 \\
\\
$\Sigma^{*+}$(1382.8) - $\Lambda^{0}$\hphantom{00} & 267.117 & 9 & 263.862 & 3.255 \\
\\
$\Sigma^{*+}$(1382.8) - $\Sigma^{0}$\hphantom{00} & 190.158 & 7 & 205.226 & 15.068 \\
& & ($\frac{13}{2}$) & 190.567 & 0.409 \\ 
  & \\ 
$\Sigma^{*}_{avg}$(1384.56) - $\Sigma_{avg}$(1193.153)\hphantom{00} & 191.407 & 7 & 205.226 & 13.816 \\
& & ($\frac{13}{2}$) & 190.567 & 0.840 \\ 
\\
$\Sigma^{*+}$(1382.8) - $\Xi^{-}$\hphantom{00} & 61.49 & 2 & 58.636 & 2.854\\
\\
$\Sigma^{*-}$(1387.2) - $\Xi^{0}$\hphantom{00} & 72.37 & 2 & 58.636 & 13.734\\
& & ($\frac{5}{2}$) & 73.259 & 0.925 \\
\\
$\Xi^{*0}$(1531.8) - n\hphantom{00} & 592.234 & 20 & 586.36 & 5.87\\
\\
$\Xi^{*-}$(1535) - $\Sigma^{-}$\hphantom{00} & 337.551 & 12 & 351.816 & 14.265\\
& & ($\frac{23}{2}$) & 337.157 & 0.394 \\
\\ 
$\Xi^{*-}$(1535) - $\Xi^{0}$\hphantom{00} & 220.17 & 7 & 205.226 & 14.944\\
& & ($\frac{15}{2}$) & 219.885 & 0.285 \\
\\ 
$\Omega^{-}$ - n\hphantom{00}& 733.884 & 25 & 732.95 & 0.934 \\
\\
$\Omega^{-}$ - $\Lambda^{0}$\hphantom{00}& 556.767 & 19 & 557.042 & 0.275 \\
\\
$\Omega^{-}$ - $\Xi^{-}$\hphantom{00}& 351.14 & 12 & 351.816 & 0.676 \\
\\
\botrule

\end{tabular} \label{ta1}}
\end{table} 
\begin{table}[h]
\tbl{The observed mass intervals between pseudoscalar and vector mesons as integral multiple of 29.318 MeV}
{\begin{tabular}{@{}lllllll@{}} \toprule
Particles & Mass Difference & integer  & $N$$\times$29.318  & Obsd - Expd  \\
& (MeV) & $N$ &  (MeV) & (MeV)\\ \colrule
k$^{*0}$(896) - $\pi^{0}$\hphantom{00} & 761.03 & 26 & 762.263 & 1.233 \\
\\
k$^{*\pm}$(891.66) - k$^{\pm}$\hphantom{00} & 397.983 & 14 & 410.452 & 12.469 \\
& & ($\frac{27}{2}$) & 395.793 & 2.19
\\
\\
k$^{*0}$(896) - $\eta$\hphantom{00} & 348.49 & 12 & 351.816 & 3.326 \\
\\
k$^{*0}$(896) - $\eta^{'}$\hphantom{00} & 61.73 & 2 & 58.636 & 3.094 \\
\\
$\rho$ - k$^{0}$\hphantom{00} & 277.852 & 10 & 293.18 & 15.328 \\
& & ($\frac{19}{2}$) & 278.521 & 0.669
\\
\\
$\omega$ - $\pi^{0}$\hphantom{00}   & 647.6734   & 22 & 644.996 & 2.677   \\
\\
$\omega$ - k$^{\pm}$\hphantom{00} &   288.823    & 10 & 293.18 & 4.207 \\
\\
$\omega$ - $\eta$\hphantom{00} &    235.14    & 8 & 234.544 & 0.596 \\ 
\\
$\omega$ - $\eta^{'}$\hphantom{00} &  175.13    & 6 & 175.908 & 0.778 \\ 
\\
$\phi$ - $\pi^{\pm}$\hphantom{00} &  879.89   & 30 & 879.54 & 0.350 \\ 
\\
$\phi$ - k$^{\pm}$\hphantom{00} &  525.783    & 18 & 525.724 & 1.94 \\
\\
$\phi$ - $\eta$\hphantom{00} & 471.95 & 16 & 469.088 & 2.862 \\
\\
$\phi$ - $\eta^{'}$\hphantom{00} & 61.68 & 2 & 58.636 & 3.044 \\
\\
 \botrule

\end{tabular} \label{ta1}}
\end{table}
\subsection{Neutral Hadrons}

The neutral hadrons are placed in the Column 1 of the Table 7 in the ascending order of their associated physical mass.  Since no bound system of top quarks has been detected, the neutral hadrons listed in Table 7 span the quark structure from the light u, d, s quarks to the heaviest charm and bottom quarks. The masses of the heavier mesons are :$m_{J/\Psi(1S)}$ = 3096.916 MeV, $m_{\Psi(2S)}$ = 3686.093 MeV, $m_{\Upsilon(1S)}$ = 9460.30 MeV. The differences amongst mesons $\pi^{0}$, $\eta$, $\eta^{'}$ and $\omega$ reported in tables 3 and 6 are reproduced here for completeness. The $\Sigma^{0}$ is the only baryon in the list. In column 2 we give the mass differences between the successive particles. The integral (Column 3) multiples of 29.318 MeV, the mass difference between a neutral pion and a muon, closest to the observed mass differences are tabulated in Column 4 and in Column 5 we give the deviation of the observed value from the predicted value. As evident from the table the agreement of the predicted values as integral multiples of pion muon mass difference with the observed values is extraordinary. The observed mass difference between $\Sigma^{0}$ and $\eta^{'}$ is 234.862 MeV. 
\begin{table}[h]
\tbl{The successive mass differences among the neutral hadrons as integral multiple of 29.318 MeV}
{\begin{tabular}{@{}lllllll@{}} \toprule
Particles & Mass Difference & integer  & $N$$\times$29.318  & Obsd - Expd  \\
& (MeV) & $N$ &   & (MeV) &  & \\ \colrule
$\pi^{0}$\hphantom{00} & 412.78 & 14 & 410.452 & 2.328 \\
\\

$\eta$ \hphantom{00}   & 234.9   & 8 & 234.5445 & 0.356   \\
\\

$\omega$ \hphantom{00} &    175.13    & 6 & 175.908 & 0.778 \\

\\
$\eta^{'}$\hphantom{00} &  234.862 & 8 & 234.544 & 0.318 \\ 

\\ 
$\Sigma^{0}$\hphantom{00} & 1904.274 & 29 & 1905.67 & 1.396 \\
\\

J/$\Psi(1S)$\hphantom{00} & 589.177 & 20 & 586.36 & 2.817 \\

\\
$\Psi(2S)$\hphantom{00} & 5773.91 & 197 & 5775.646 & 1.736 \\
\\
$\Upsilon(1S)$\hphantom{00} &  &  &  &  \\
\botrule
\end{tabular} \label{ta1}}
\end{table}
This value is very close to 234.544 MeV obtained on multiplication of 29.318 MeV by integer 8. The difference between the two values being only 0.318 MeV. Similarly 1904.274 MeV, the observed difference between the mass of J/$\Psi(1S)$ and   that of $\Sigma^{0}$ deviates from 29 multiples of 29.318 MeV i.e. 1905.67 MeV by only 1.396 MeV. The J/$\Psi(1S)$ is the lowest-lying member of the charm-anticharm meson series and $\Psi(2S)$ is its first excited state. From Table 7 it can be seen that the mass difference between $\Psi(2S)$ and J/$\Psi(1S)$ is close integral multiple of the mass difference between a neutral pion and muon. Similarly $\Upsilon(1S)$ is the first member of the bottom-antibottom meson series. From table 13, it is seen that the observed mass difference between the $\Upsilon(1S)$ and J/$\Psi(1S)$ is 6363.384 which is very close to 217 multiples of 29.318 MeV i.e. 6362.006. The difference between the two values being only 1.378 MeV. The fact that the mass differences of the lightest hadron $\pi^{0}$ with the heaviest mesons J/$\Psi(1S)$, $\Psi(2S)$  and $\Upsilon(1S)$ are close integral multiples of the mass difference between a neutral pion and muon indicates that the 29.318 MeV mass systematics is valid over the wider range of the mass spectrum. 
\begin{table}[h]
\tbl{Mass intervals of neutral hadrons with respect to $\pi^{0}$ as integral multiple of 29.318 MeV}
{\begin{tabular}{@{}lllllll@{}} \toprule
$\pi^{0}$ & Mass Difference & integer  & $N$$\times$29.318  & Obsd - Expd  \\
& (MeV) & $N$ &  (MeV) & (MeV)\\ \colrule

$\omega$\hphantom{00}   & 647.6734   & 22 & 644.996 & 2.677   \\
\\
$\eta^{'}$\hphantom{00} & 822.81 & 28 & 820.904 & 1.906 \\
\\
$\Sigma^{0}$\hphantom{00} & 1057.6654 & 36 & 1055.448 & 2.212 \\ 
\\
J/$\Psi(1S)$\hphantom{00} &  2961.9394 & 101 & 2961.118 & 0.821 \\ 
\\
$\Psi(2S)$\hphantom{00} &  3551.1164    & 121 & 3547.478 & 3.6384 \\ 
\\
$\Upsilon(1S)$\hphantom{00} &  9325.3224    & 318 & 9323.124 & 2.199 \\ \botrule

\end{tabular} \label{ta1}}
\end{table}
In tables 8 to 13 we detail the analysis for the mass intervals of each neutral hadron with respect to the remaining hadrons. The mass intervals for each neutral hadron already listed in Table 7 have been excluded in the analysis detailed in 8-13. From Tables 1 to 13 it is clear that the hadron mass intervals fall into two classes 1) those integral multiples of the mass difference between a neutral pion and a muon and 2) those in which the difference between the observed and predicted values are large but turn out to be exact half integral multiples of 29.318 MeV.   
\begin{table}[h]
\tbl{Mass intervals of neutral hadrons with respect to $\eta$ as integral multiple of 29.318 MeV}
{\begin{tabular}{@{}lllllll@{}} \toprule
$\eta$ & Mass Difference & integer  & $N$$\times$29.318  & Obsd - Expd  \\
& (MeV) & $N$ &  (MeV) & (MeV)\\ \colrule

$\eta^{'}$\hphantom{00} & 410.27    & 14 & 410.452 & 0.182 \\
\\
$\Sigma^{0}$\hphantom{00} & 644.892 & 22 & 644.996 & 0.104 \\ 
\\
J/$\Psi(1S)$\hphantom{00} &  2549.116 & 87 & 2550.666 & 1.55 \\ 
\\
$\Psi(2S)$\hphantom{00} &  3138.583 & 107 & 3137.062 & 1.521 \\ 
\\
$\Upsilon(1S)$\hphantom{00} & 8912.55 & 304 & 8912.612 & 0.122 \\
\botrule
\end{tabular} \label{ta1}}
\end{table}

\begin{table}[h]
\tbl{Mass intervals of neutral hadrons with respect to $\omega$ as integral multiple of 29.318 MeV}
{\begin{tabular}{@{}lllllll@{}} \toprule
$\omega$ & Mass Difference & integer  & $N$$\times$29.318  & Obsd - Expd  \\
& (MeV) & $N$ &  (MeV) & (MeV)\\ \colrule

$\Sigma^{0}$\hphantom{00} & 409.992 & 14 & 410.452 & 0.460 \\ 
\\
J/$\Psi(1S)$\hphantom{00} &  2314.266 & 79 & 2316.122 & 1.856 \\ 
\\
$\Psi(2S)$\hphantom{00} &  2903.443 & 99 & 2902.482 & 0.961 \\ 
\\
$\Upsilon(1S)$\hphantom{00} &  8677.65  & 296 & 8678.128 & 0.478\\
\botrule

\end{tabular} \label{ta1}}
\end{table}

\begin{table}[h]
\tbl{Mass intervals of neutral hadrons with respect to $\eta^{'}$ as integral multiple of 29.318 MeV}
{\begin{tabular}{@{}lllllll@{}} \toprule
$\eta^{'}$ & Mass Difference & integer  & $N$$\times$29.318  & Obsd - Expd  \\
& (MeV) & $N$ &  (MeV) & (MeV)\\ \colrule

J/$\Psi(1S)$\hphantom{00} &  2139.136 & 73 & 2140.214 & 1.078 \\ 
\\
$\Psi(2S)$\hphantom{00} &  2728.313 & 93 & 2726.574 & 1.739 \\ 
\\
$\Upsilon(1S)$\hphantom{00} & 8502.22    & 290 & 8502.22 & 0.300 \\
\botrule

\end{tabular} \label{ta1}}
\end{table}

\begin{table}[h]
\tbl{Mass intervals of neutral hadrons with respect to $\Sigma^{0}$ as integral multiple of 29.318 MeV}
{\begin{tabular}{@{}lllllll@{}} \toprule
$\Sigma^{0}$ & Mass Difference & integer  & $N$$\times$29.318  & Obsd - Expd  \\
& (MeV) & $N$ &  (MeV) & (MeV)\\ \colrule

$\Psi(2S)$\hphantom{00} &  2493.451 & 85 & 2492.03 & 1.421 \\ 
\\
$\Upsilon(1S)$\hphantom{00} & 8267.65   & 282 & 8267.676 & 0.026 
\\
\botrule

\end{tabular} \label{ta1}}
\end{table}
\begin{table}[h]
\tbl{Mass intervals of neutral hadrons with respect to J/$\Psi(1S)$ as integral multiple of 29.318 MeV}
{\begin{tabular}{@{}lllllll@{}} \toprule
J/$\Psi(1S)$ & Mass Difference & integer  & $N$$\times$29.318  & Obsd - Expd  \\
& (MeV) & $N$ &  (MeV) & (MeV)\\ \colrule

$\Upsilon(1S)$\hphantom{00} &  6363.384    & 217 & 6362.006 & 1.378 \\
\botrule

\end{tabular} \label{ta1}}
\end{table}
\section{Discussion}
The inability of the standard model to explain the observed spectra of mass of elementary particle has led to the development of models and theories for particle masses based on direct analysis of the elementary particle data to understand different intricacies of the mass spectrum. The recognition of the pionic mass intervals among the particles\cite{Mac Gregor2} combined with the fact that ratios of life times of elementary particles scale as powers of $\alpha$, the fine structue constant, indicated the fundamental role of $\alpha$ in the generation of mass of elementary particles and that its domain is not restricted to leptons only, but also extends to include hadrons\cite{Mac Gregor4}. The statistical analysis of masses of elementary particles when grouped according to their quark composition and $J^{PC}$ quantum numbers reveals that the masses are quantized as integral multiples of a mass unit of about 35 MeV. This 35 MeV mass quantization combined with the analysis of the distribution of particle lifetimes as a function of mass reveals a shell structure of hadrons\cite{Palazzi}. A standing wave model assuming the particles to be held together in a cubic nuclear lattice has been proposed to explain empirical fact that meson and baryon masses are integral multiples of the neutral pion\cite{Koschmieder}. From the preceeding discussion, it is clear that the understanding of any pattern in a physical system is a driving force for the development of a fundamental theory. The development proceeds from the experimental measurements of the phenomena, first to a recognition of regularities amongst the measurements, then to the physical insight which gives some understanding of these regularities and finally to a fundamental theory, which allows the totality of the phenomena to be understood from a few general principles\cite{Hall}. The failure to understand the regularities within a reasonable data lays the foundation for the neccessity of the underlying theory.
\newline
The above mentioned studies are among many other attempts dedicated to pin down the fundamental relation of least massive and most stable particles, pion and muon, to the elementary particle mass spectrum pointed by their presence in the decay products of almost all the heavier elemenatry particle mass states\cite{Nambu}$^{-}$\cite{Akers}. The mass intervals among elementary particles have been shown to be integral multiples of the pion mass\cite {Mac Gregor2} and some sequances of particles are also repoted with muonic mass intervals\cite{Sternheimer1}. The baryon and meoson masses are reported to be integral/half integral multiples of the pion mass\cite{Sternheimer3}$^{-}$\cite{Koschmieder}. The mass unit of 35 MeV equal to one-third of muon mass and one-fourth of the pion mass serves as basic unit for the quantization of the elementary particle masses\cite{Palazzi}. On the other hand we look at the elementary particle mass data from a different prespective and reveal the fundamental nautre of mass difference between a neutral pion and a muon to the elementary particle mass spectrum. The electron and muon mass are the first two levels in the mass spectrum. Indeed the muon is treated to be an excited state of the electron\cite{Mac Gregor4}$^{,}$ \cite{Barut}$^{,}$ \cite{Rodrigues}. However, the mass of electron being much smaller than the muon mass, the mass interval between the two is obscured by the large mass of muon and is thus equal to muon mass itself. Hence we regard 29.318 MeV, the mass interval between neutral pion (a hadron) and muon (a lepton) to be the first excitation within elementary particle mass spectrum. We investigate whether the mass excitations among the elementary particles are related to the first excitation. 
\newline
We have satisfactorily shown that mass differences among the elementary particles in general when arranged in the ascending order of mass have a striking tendency to be integral/half integral multiples of 29.318 MeV\cite{Shah}. Same is shown to be true for the mass differences between any of the baryons and mass interval between the unstable charged leptons. From the present study, it follows that the "fine structure" of mass i.e. mass differences betweend different isospin members of a given hadron multiplet are close integral multiples of the mass difference between a neutral pion and a muon. This is also true for the mass separation of the lightest and heaviest member. The mass intervals between the members of different SU(3) multiplets, i.e. between octet and decuplet baryons and between the pseudoscalar and vector mesons are also integral multiples of the mass unit of 29.318 MeV. The 29.318 MeV discretness of the mass intervals is applicable to elementary particles irrespective of their classification into leptons, mesons and baryons on the basis of their structure, associated quantum numbers and type of interaction. It covers a wide range of mass spectrum, a fact borne out by the precise agreement of the observed and predicted mass intervals for the heavy charm-anticharm and bottom-antibottom mesons. Although the present findings are not in line with the current theories, yet the large scale agreement between the observed mass differences and those calculated as integral multiples of the mass difference between a neutral pion and a muon clearly rules out it to be a mere coincidence. The important intriguing result of the present study is that the mass difference between the lightest composite particle pion and the second lightest lepton muon interlinks the masses of both leptons and hadrons. Clearly the occurence of elementary particle states is not random but seems to follow a definite order such that the mass excitations from one particle to other are always in units of 29.318 MeV i.e. the mass difference between a neutral pion and a muon. 
\section{Conclusion}
The mass differences among the baryon octet members have been found to be close integral multiples of the mass difference between a neutral pion and a muon. This also holds true for the average mass separation among the successive members of the baryon decuplet, for the mass intervals among the pseudoscalar meson nonet and among the vector meson nonet members. We also pointed out mass intervals among neutral hadrons to be integral multiples of mass difference between a neutral pion and a muon. Our study reveals that inter particle mass excitations are quantized as integral multiples of 29.318 MeV thereby indicating the fundamental nature of this mass quantum for the elementary particle mass spectrum.

%\section*{Acknowledgments}
%The authors are highly thankful to Paolo Palazzi, Amjad Hussain Shah Gilani and David Akers for their helpful comments and suggestions on the contents of the paper. 

%This section should come before the References. Dedications and funding 

%\section*{References}


\begin{thebibliography}{00}
\bibitem{Christianto} V. Christianto and F. Smarandache, {\text Progress in Physics, 4}, 112 (2007)
\bibitem{Fritzsch} H. Fritzsch, {\text arxiv:0802.0099, hep-ph/0201198}
\bibitem{Okubo} S. Okubo, {\text Prog. Th. Phys. 27}, 949 (1962)
\bibitem{Coleman} S. Coleman and S. L. Glashow, {\text Phys.  Review Lett. 6}, (1961)
\bibitem{Mac Gregor3} M. H. Mac Gregor, {\text IL Nuovo Cimento. A103, 983 (1990)}
\bibitem{Shah} G. N. Shah and T. A. Mir, {\text hep-ph/0702140}, {\text Mod. Phys. Lett. A23, 53 (2008)} 
\bibitem{Nambu} Y. Nambu, {\text Prog. Th. Phys. 7}, 595 (1952)
\bibitem{Sternheimer1} R. M. Sternheimer, {\text Phys.  Review Lett. 10, 7 (1963), 13, 1 (1964), 13, 10 (1964)}
\bibitem{Sternheimer2} R. M. Sternheimer, {\text Phys. Review. B2, 138 (1965), 143, 4 (1966)}
\bibitem{Sternheimer3} R. M. Sternheimer, {\text Phys.  Review, 143, 1262 (1966), 5, 170 (1968)}
\bibitem{Tangherlini} F. R. Tangherlini, {\text Prog. Th. Phys. 58}, 2002 (1977)
\bibitem{Sidharth} B. G. Sidharth, {\text physics/0309037}, {\text physics/0306010}
\bibitem{Mac Gregor2} M. H. Mac Gregor, {\text IL Nuovo Cimento. A58, 159 (1980)}
\bibitem{Mac Gregor4} M. H. Mac Gregor, {\text Int. J. Mod. Phys. A20, 719 (2005)}
\bibitem{Palazzi} P. Palazzi, {\text Int. J. Mod. Phys. A22}, 546 (2007), (URL: http://www.particlez.org)
\bibitem{Koschmieder} E. L. Koschmieder, {\text physics/0602037}, {\text physics/0002179}
\bibitem{Frosch} R. Frosch, {\text Nuovo Cimento. A104, 913 (1991)}
\bibitem{Barut} A. O. Barut, {\text Phys. Lett. B73}, 310 (1978), {\text Phys. Rev. Lett. 42}, 1251 (1979)
\bibitem{Rodrigues} W. A. Rodrigues and J. Vaz, {\text hep-th/9607231}
\bibitem{Akers} D. Akers, {\text hep-ph/0310239}, {\text hep-ph/0311031}, {\text hep-ph/0303139}, {\text hep-ph/0303025}
\bibitem{Yao} W.- M. Yao et al, {\text (Particle Data Group), J. Phys. G. 33, 1}, (2006) (URL: http://pdg.lbl.gov)
\bibitem{Perkins} D. H. Perkins, {\text Introduction to High Energy Physics}, pp. 126-129, 4th Edn. (Cambridge University Press, 2000)
\bibitem{Oh} Y. Oh, H. Kim and S. H. Lee, {\text hep-ph/0310117}
\bibitem{Hall} L. J. Hall, {\text hep-ph/9303217}

\end{thebibliography}
\end{document}